\documentclass[aps,prc,twocolumn,superscriptaddress,showpacs,nofootinbib,floatfix]{revtex4}

\usepackage{graphicx} \usepackage{dcolumn} \usepackage{bm}
\newcommand{\beq}{\begin{equation}}\newcommand{\eeq}[1]{\label{#1}
\end{equation}}\newcommand{\beqar}{\begin{eqnarray}}\newcommand{\eeqar}[1]{\label{#1}
\end{eqnarray}}\newcommand{\bmath}{\begin{displaymath}}\newcommand{\emath}{\end{displaymath}}\newcommand{\bitem}{\begin{itemize}}\newcommand{\eitem}{\end{itemize}}
 
\begin{document}

\title{\Large \bf Soft Open Charm Production in Heavy-Ion Collisions.}

\newcommand{\mcgill}{McGill, University, Montreal, Canada, H3A 2T8}

\newcommand{\columbia}{Physics Department,
Columbia University, New York, N.Y. 10027}

\affiliation{\mcgill}
\affiliation{\columbia}

\author{~V.~Topor~Pop} \affiliation{\mcgill}
\author{~J.~Barrette} \affiliation{\mcgill}
\author{~M.~Gyulassy} \affiliation{\columbia}

\date{February 23, 2009}

\begin{abstract}

Effects of strong longitudinal color 
electric fields (SCF) on the open charm production in nucleus-nucleus
 ({\it A} + {\it A}) collisions at 200A GeV  are investigated 
within the framework of 
the {\small HIJING/B\=B v2.0} model.
A three fold increase of the effective string tension
due to {\em in medium effects} in {\it A} + {\it A} collisions,
results in a sizeable ($ \approx $ 60-70 \%) enhancement 
of the total charm production cross sections
($\sigma^{\rm NN}_{{\it c}\,\bar{{\it c}}}$). 
The nuclear modification factors show a
suppression at moderate transverse momentum (${\rm p}_{\rm T}$)
consistent with RHIC data. 
At Large Hadron Collider energies 
the model predicts an increase of $\sigma^{\rm NN}_{{\it c}\,\bar{{\it c}}}$ 
by approximately an order of magnitude.

\end{abstract}


\maketitle





The phase transition from hadronic degrees of freedom to
partonic degrees of freedom in ultra-relativistic nuclear collisions 
is a central focus of experiments at the Relativistic 
Heavy Ion Collider (RHIC).
Heavy-flavor quarks are an ideal probe to study early dynamics 
in these nuclear collisions.
Several theoretical studies predict  
\cite{wang_prl92,shuryak_prl92,geiger_prc93} 
a substantial enhancement of open 
charm production associated to plasma formation of the deconfined 
parton matter relative to the case of a purely hadronic scenario 
without plasma formation.
A recent analysis shows that the dynamics of heavy-quarks at 
RHIC are dominated by partonic or ``pre-hadronic'' interactions 
in the strongly coupled plasma (sQGP) stage and can neither 
be modeled by ``hadronic interactions'' nor described appropriately
by color screening alone \cite{cassing_08}.
Therefore, these quarks are key observables in the study   
of thermalization of the initially created hot nuclear matter
 \cite{nuxu_07}.

A review of heavy-flavor production in heavy-ion collisions 
has been recently published \cite{vogt08_1}.
Direct reconstructed $D^{0}$ mesons 
via hadronic channel ($D^0\,\rightarrow K \pi $) 
in {\it d} + Au \cite{star_05}, Cu + Cu \cite{star_08_1}  and 
Au + Au \cite{star_08_2} collisions have been measured.    
Due to the difficulty to reconstruct {\it D}-mesons hadronic decay vertex,
both STAR and PHENIX have studied open charm indirectly
via semileptonic decay to non-photonic electrons (NPE) or muons
\cite{star_05,star_08_2,star_07,yzhang_08,phenix_05,phenix_06_1,phenix_06_2,phenix_07}. Theory predict that charm quarks are produced by initial
gluon fusion \cite{Lin95} and
their production rates are expected to be well described 
by perturbative Quantum Cromodynamics (pQCD)
at Fixed Order plus Next to-leading Logarithms (FONLL) \cite{cacciari_05}.
Total charm cross sections reported by both experiments 
are however only compatible with the upper limit 
of the FONLL predictions.
In addition, the data indicate a suppression 
as large as that of light quarks \cite{star_07},\cite{phenix_07}
while, due to their large mass and to {\em the dead cone effect} 
charm quarks are predicted to loose 
less energy than light quarks by gluon radiation in the medium 
\cite{kharzeev_01}.
 
Recent model calculations based on in-medium charm 
resonances/diffusion or collisional dissociation 
\cite{Rapp_06,Vitev_07,Moore_05},
radiative energy loss via few hard scatterings \cite{Magdalena_06} 
or radiative energy loss via multiple soft collisions 
\cite{Armesto_06}, have been applied to describe the non-photonic 
electrons (NPE) spectra. They all predict less suppression than 
that observed in experiments. On the other hand,
a good description of the nuclear modification factor
(NMF), ${\rm R}_{\rm AA}^{\rm npe}({\rm p}_{\rm T})$, was obtained in 
non-perturbative time-dependent heavy-quark diffusion in the 
Quark-Gluon Plasma (QGP) \cite{rapp_prl08}.

In previous papers \cite{prc72_top05,prc75_top07} we have shown 
that the dynamics of strangeness production 
deviates considerably from calculations based on Schwinger-like
estimates for homogeneous and constant color fields \cite{schwinger} 
and point to the 
contribution of fluctuations of transient strong 
color fields (SCF).
These fields are similar with those which could appear in a 
 ``{\em glasma}'' \cite{mclerran_08} at initial stage of the collisions.
In a scenario with QGP phase
transitions the typical field strength of SCF at RHIC energies 
was predicted to be about 5-12 GeV/fm \cite{csernai_npa02}.
Recently Schwinger mechanism has been revisited \cite{cohen_jul08} and
pair production in time-dependent
electric fields has been studied \cite{gies_jul08}.
It is concluded that particles with large momentum are likely to have 
been created earlier than particle with small momentum and 
for very short temporal widths 
($\Delta \tau \approx 10 {\rm t}_{\rm c}$, where the Compton time 
${\rm t}_{\rm c}=1/{\rm m}_{\rm c}$) the  Schwinger formula strongly
underestimates the reachable particle number density.


In this paper we extend our study in the framework of  
{\small HIJING/B\=B v2.0} model \cite{prc75_top07}
to open charm productions.  
We explore dynamical effects associated with
long range coherent fields (i.e strong color fields, SCF),
including baryon junctions and loops \cite{prc72_top05}, with emphasis 
on the novel open charm observables measured at RHIC
in {\it p}+{\it p} and heavy-ion collisions.
Using this model we analyze the enhancement of total charm production
at 200A GeV energy.

For a uniform chromoelectric flux tube with field ({\it E}) 
the pair production rate \cite{Biro:1984cf, Gyulassy:1986jq, cohen_jul08}  
per unit volume for a heavy quark is given by:
\begin{equation}
\Gamma =\frac{\kappa^2}{4 \pi^3} 
{\text {exp}}\left(-\frac{\pi\,m_{Q}^2}{\kappa}\right)
\end{equation}
where for $Q=c$ or $b$ , $m_{\rm Q}=1.27,\;{\rm or}\; 4.16$ 
GeV (with $\pm 1\%$ 
uncertainty \cite{Steinhauser:2008pm}) 
Note that $\kappa=|eE|_{eff}= \sqrt{C_2(A)/C_2(F)}\, \kappa_0$ 
is the effective string tension
in terms of the vacuum string tension $\kappa_0 \approx 1 $ GeV/fm
and $C_2(A)$, $C_2(F)$ are the second order Casimir operators 
(see Ref. \cite{Gyulassy:1986jq}).
In a nuclear collisions, the local longitudinal 
field strength increases with the square root of the 
number of color exchanges proportional to 
the number of binary collisions per unit area
($\kappa(x_\perp,b)\propto \sqrt{T_AA(x_\perp,b)}$),
 where for a given impact parameter
$b$ and transverse coordinate $x_\perp$, $T_AA \propto A^{2/3}$ 
is the Glauber A+A 
binary collision distribution. Therefore, the effective string tension 
$\kappa \propto A^{1/3}$.

A measurable rate for spontaneous pair production requires 
``{\em strong chromo electric fields}'', such 
that $\kappa/m_{\rm Q}^2\,\,>$ 1 {\em at least some of the time}.
On the average, longitudinal electric field ``string'' models 
predict for heavier flavor a very suppressed production rate per unit
volume $\gamma_{Q}$ via the well known Schwinger formula \cite{schwinger}, 
since
\begin{equation}
\gamma_{Q\bar{Q}} = \frac{\Gamma_{Q\bar{Q}}}{\Gamma_{q\bar{q}}} =
{\text {exp}} \left(-\frac{\pi(m_{Q}^2-m_q^2)}{\kappa_0} \right)
\ll 1
\end{equation}
for $Q=c$ and $q=u,d$.
For a color rope on the other hand,
if the  {\em average} string tension value ($<\kappa>$) 
increases from 1.0 GeV/fm to 3.0 GeV/fm,
the rate $\Gamma$ for charm pairs to tunnel through the longitudinal field
increases from $\approx 1.4\,\cdot 10^{-12}$ to 
$\approx 3.5\,\cdot 10^{-4}$ fm$^{-4}$,
and this can lead to  a net ``soft'' tunneling production
comparable to the initial ``hard'' FONLL pQCD production.

The conventional hard pQCD  mechanism, mainly 
{\em gluon fusion} \cite{wang_prl92}, is calculated via the {\small PYTHIA}
subroutines in {\small HIJING/B\=B v2.0}. 
The advantage of {\small HIJING} over {\small PYTHIA}
is the ability to include novel SCF color rope effects 
that arise from longitudinal fields amplified by the random walk 
in color space of the high x valence partons in 
{\it A}+{\it A} collisions. 
This random walk could induce   
a very broad fluctuation spectrum of the effective string tension.
Thus, if the average or mean $<\kappa> = n\,\kappa_{0}$, then the typical 
fluctuation is of order $1/\sqrt{n}$ which is large because 
$n \approx 6$ for Au nuclei. 
A Poisson fluctuation of effective $\kappa$ about the mean, gives a 
strong bias toward less probable but larger 
than the average value $<\kappa>$. This is amplified for heavy 
quarks. 
Here we do not investigate in details such fluctuations, but we will
estimate the effects of a larger effective value $\kappa > $ 3 GeV/fm
on the enhancement of $\sigma^{\rm NN}_{{\it c}\,\bar{{\it c}}}$.

\begin{figure} [hb!]
\centering
\includegraphics[width=\linewidth,height = 4.5 cm]{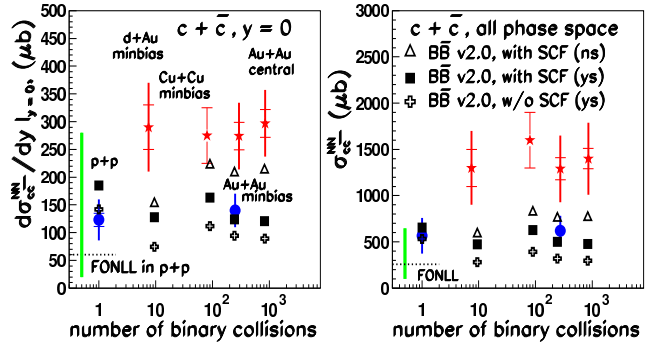}
\vskip 0.01cm\caption[ccbar cross sections at y=0, ally] {\small (Color online) Comparison of {\small HIJING/B\=B v2.0} predictions for
mid-rapidity (left panel) and all phase space (right panel) 
charm cross sections per nucleon-nucleon collisions as a function 
of ${\rm }N_{\rm bin}$ in ({\it d}){\it A}+{\it A} collisions.
The symbols are the results with (filled squares) and 
without (open crosses) SCF effects. Both include quenching and
shadowing (ys) effects. The open triangles are the results 
with SCF effects but no shadowing (ns).
The values of FONLL predictions are shown as a dotted line. 
The band at the left mark the FONLL uncertainties \cite{cacciari_05}.  
The data are from STAR (stars)
\cite{star_05},\cite{star_08_1},\cite{star_08_2},\cite{yzhang_08}
and PHENIX (solid circles) \cite{phenix_05},\cite{phenix_06_2}.
Statistical and systematical error bars are shown.
\label{fig:fig1}
}
\end{figure}

Both STAR and PHENIX experiments have measured charm production cross 
sections in several collision systems.
Figure~\ref{fig:fig1} shows the measured total charm production cross 
sections at mid-rapidity, 
$d\,\sigma^{\rm NN}_{{\it c}\,\bar{{\it c}}}/dy$ (left
panel) and in all phase space, 
$\sigma^{\rm NN}_{{\it c}\,\bar{{\it c}}}$ (right panel).
The data from both experiments seems to indicate a scaling 
with number of binary collisions (${\rm N}_{\rm bin}$), 
as expected because of the high mass of charm pairs produced in initial
nucleon-nucleon collisions \cite{Lin95}. 
However, there is still an unresolved discrepancy of the order of
a factor of two between STAR and PHENIX data.

The predictions of {\small HIJING/B\=B v2.0} model without SCF (open crosses)
and including SCF effects (filled squares) are shown in the figure.
For completeness the results with SCF but no gluon 
shadowing effects (open triangles) are also included. 
However, in this scenario multiplicities at
mid-rapidity are strongly overestimated \cite{topor_prc03}.
The main parameters used in the calculations are given in 
Table II of reference \cite{prc75_top07}, 
and corresponds to 
strengths of strong color (electric) field dependent on collision system
($\kappa$ = 1.5; 2.0; 3.0 GeV/fm for {\it p} + {\it p}, {\it d} + Au, and 
{\it A} + {\it A} collisions respectively).
In our calculations we estimate the total open charm 
 production ({\it c} + $\bar{\it c}$) cross section considering 
the 12 lightest {\it D}-mesons (${\rm D}^{0}$, ${\bar {\rm D}}^{0}$,
${\rm D}^{0*}$, ${\bar {\rm D}}^{0*}$,
${\rm D}^{+}$, ${\bar {\rm D}}^{+}$,  ${\rm D}^{+*}$, ${\bar {\rm D}}^{+*}$,
${\rm D}_{\rm s}$, ${\bar {\rm D}}_{\rm s}$,
${\rm D}^{*}_{\rm s}$, ${\bar {\rm D}}^{*}_{\rm s}$), and the hyperons 
$\Lambda_{\rm c}$ and $\bar{\Lambda_{\rm c}}$.
The contribution of higher mass charm hyperons is negligible.
For calculations which take into consideration SCF effects (filled
squares) we obtain an increase of $60-70 \,\%$ in comparison 
with a scenario without SCF effects (open crosses). 
These results describe well the PHENIX data within  statistical
and systematical errors and are close to the upper limit of  
uncertainty band of the pQCD FNOLL predictions \cite{cacciari_05}.
Our calculations also show that the scaling with ${\rm N}_{\rm bin}$
is only approximately satisfied, the reason being 
an interplay between the mass dependent SCF and 
shadowing effects, which act in opposite directions.
In fact, we calculate that only $60\, \%$ of total open charm production 
({\it c} + $\bar{\it c}$) comes from partons embedded within the target 
and projectile.

The study of open charm production in {\it d} + Au collisions 
allow to separate ``cold nuclear matter'' effects.
The initial production
of {\it c} $\bar{\it c}$ pairs by gluon fusion might be suppressed 
due to gluon shadowing.
We recall that shadowing is a depletion of the low-momentum parton
distribution in a nucleon embedded in a nucleus compared to a free
nucleon; this leads to a lowering in the (scaled) 
{\it c} + $\bar {\it c}$ production relative to {\it p} + {\it p}
collisions.
The shadowing in the regular HIJING parameterization \cite{hij_94}
implemented also in our model seems to be too strong \cite{wang_pl527_02}.
There is a considerable uncertainty 
(up to a factor of 3) in the amount of shadowing
predicted at RHIC energies by the different models with HIJING 
predicting the strongest effect \cite{cms_enteria_08}.
This could explain 
why the results for scaled cross sections in {\it d} + Au collisions 
are smaller than those obtained for {\it p} + {\it p}
collisions (see Fig.~\ref{fig:fig1} left panel).

We study if we can find scenarios that would give larger enhancement
of total cross sections for open charm production, than  
those reported in Fig.~\ref{fig:fig1} (filled squares), 
and that would be consistent with the STAR data.
The random walk in color space of heavy quark could induce a broad spectrum 
of the effective string tension.
Therefore, we study the effects of    
a further increase of mean value of the string tension from
3.0 GeV/fm to 5.0 GeV/fm on ${\it c}\,\bar{{\it c}}$ pair production.
This results in only a modest increase of scaled 
cross sections, $\sigma^{\rm sNN}_{{\it c}\,\bar{{\it c}}} =
\sigma^{\rm AA}_{{\it c}\,\bar{{\it c}}}/{\rm N}_{\rm bin}$, 
by approximately $20\,\%$ for central collisions.
For values between 5 - 10 GeV/fm a saturation  
sets in, as an effect of energy and momentum conservation constraints.
In our model the multiplicative factor that accounts for 
next to leading order corrections \cite{eskola_03} in calculations of 
hard or semihard parton
scattering processes via pQCD is set to K = 2. 
Increasing this factor to K = 3.5, as suggested 
in reference \cite{eskola_03} results in an increase of 
$\sigma^{\rm sNN}_{{\it c}\,\bar{{\it c}}}$ by approximately $70 \%$
in central Au + Au collisions, but also overpredicts by $40\,\%$ 
the total charged particle production at
mid-rapidity (N$_{\rm ch}$ (y=0)).
Therefore, we conclude that the large charm cross sections obtained by
the STAR collaboration cannot be explained within our phenomenology.

\begin{figure} [h!]
\centering
\includegraphics[width=0.5\linewidth,height=5.0 cm]{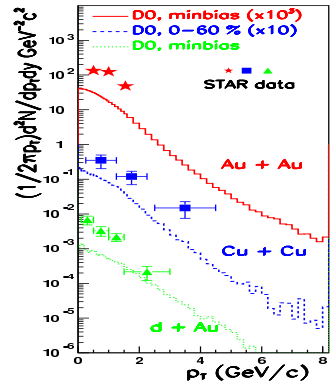}
\vskip 0.01cm\caption[pt spectra for D0, minimum-bias,central] {\small 
(Color online) Comparison of {\small HIJING/B\=B v2.0} predictions for
${\rm p}_{\rm T}$ distribution of invariant yields for reconstructed
$D^0$ in minimum-bias ({\it d}){\it A}+{\it A}  collisions.
For clarity the results for Cu + Cu (preliminary) and Au + Au are multiplied 
by 10 and $10^{3}$ respectively. The data are from STAR
\cite{star_05,star_08_1,star_08_2}. 
Only statistical error bars are shown.
\label{fig:fig2}
}
\end{figure}

The ${\it D}^0$-mesons spectra are sensitive to 
the dynamics of produced charm particles. 
In Fig.~\ref{fig:fig2} we present the calculated 
$D^0$-mesons spectra for systems where data are available \cite{star_05},\cite{star_08_1},\cite{star_08_2}.
In all cases, the calculated yield is much smaller than
the STAR data, consistent with the results shown in Fig.~\ref{fig:fig1}.
The calculated spectra show little shoulder at low ${\rm p}_{\rm T}$ 
indicating   small radial flow of ${\it D}^0$-mesons consistent with
the results of STAR \cite{yzhang_08}. 

\begin{figure} [h!]
\centering
\includegraphics[width=0.5\linewidth,height=4.0cm]{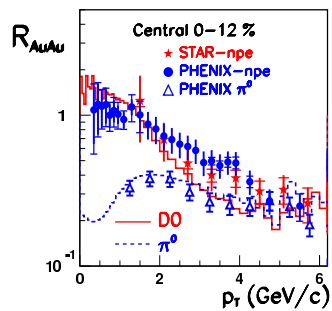}
\vskip 0.01cm\caption[R_AA of D0 for Au +Au relative to pp] {\small (Color online) Comparison of {\small HIJING/B\=B v2.0} predictions of
nuclear modification factor
${\rm R}_{\rm AA}({\rm p}_{\rm T})$ for $D^0$ and $\pi^0$ mesons in central
(0-12 \%) Au + Au collisions.
Data  from STAR (stars) \cite{star_07} and PHENIX (circles) \cite{phenix_07}
are NMF for  non photonic electrons, 
${\rm R}_{\rm AA}^{\rm npe}({\rm p}_{\rm T})$.
The data for $\pi^0$ meson are from PHENIX \cite{phenix_03}.
Error bars show the statistical and systematic uncertainties.  
\label{fig:fig3}}
\end{figure}

Figure \ref{fig:fig3} shows our predictions for 
the Nuclear Modification factor (NMF), 
${\rm R}_{\rm AA}({\rm p}_{\rm T})$ for $D^0$ and $\pi^0$ mesons.
Data (filled symbols) are NMF for non photonic
electrons, ${\rm R}_{\rm AA}^{\rm npe}({\rm p}_{\rm T})$ 
\cite{star_07},\cite{phenix_07}.
The data for $\pi^0$ meson (open symbols) are from reference 
\cite{phenix_03}.
Note, that non photonic electrons include also electrons  
from bottom (b) production (B $\rightarrow$ lX) 
and the yields of $D^0$ mesons could 
be affected by the decay (B $\rightarrow$ D). 
For central (0-10 \%) Au + Au
collisions we calculate a scaled total cross section for 
bottom production with (without) SCF of  
$\sigma^{\rm sNN}_{{\it b}\,\bar{{\it b}}}$ = 17.8 $\mu$b
($\sigma^{\rm sNN}_{{\it b}\,\bar{{\it b}}}$ = 0.86 $\mu$b).
These values are
few orders of magnitude lower than $\sigma^{\rm sNN}_{{\it c}\,\bar{{\it c}}}$
and this contribution is estimated to be 
negligible for ${\rm p}_{\rm T}\,< 6.0 $ GeV/c.

In our calculations for low 
${\rm p}_{\rm T}$ ($0\,<\,{\rm p}_{\rm T}\,< 2.5 $ GeV/c), 
non-perturbative production mechanism via SCF  
results in a split between $D^0$ and $\pi^0$ mesons.
The charged and $\pi^0$ mesons are suppressed due to
conservation of energy \cite{prc72_top05}. 
The yields of the ${\it D}^0$-mesons are enhanced due to an increase 
of $c\,\bar{c}$ pair production rate (see Eq. 1).
In central ($0-10\,\%$) Au + Au collisions a suppression 
at moderate $p_{\rm T}$ ($4\, <\, {\rm p}_{\rm T}\, <\, 6$ GeV/c) 
as large as that of light quarks is observed 
in contrast to previous theoretical studies \cite{kharzeev_01},
\cite{Magdalena_06,Rapp_06,Armesto_06,wicks_npa07}.
Our model predicts a suppression consistent with the data.
We can interpret this results as experimental evidence for 
``{\em in-medium mass modification}'' of charm quark, due to 
possible induced chiral symmetry restoration \cite{kharzeev_05}.
An in-medium mass modification has also been predicted 
near the phase boundary (i.e. at lower energy) in  \cite{tolos_06}.
In contrast statistical hadronization model \cite{andronic_08} 
predicts no medium  effects at top RHIC energy.

We performed calculations at the much higher 
Large Hadrons Collider (LHC) energy 
using parameters from reference \cite{top_jpg08}, 
i.e $\kappa$ = 2.0; 5.0 GeV/fm for {\it p} + {\it p} 
and central (0-10 \%) Pb + Pb collisions respectively. 
The predicted charm production cross section
is approximately an order of magnitude larger than at RHIC energy.
We obtain $\sigma^{\rm NN}_{{\it c}\,\bar{{\it c}}}$ = 6.4 mb 
in p+p collisions and a (scaled) cross section 
$\sigma^{\rm sNN}_{{\it c}\,\bar{{\it c}}}$ = 2.8 mb
for central Pb + Pb collisions (${\rm N}_{\rm bin}$ = 960
and N$_{\rm ch}$ (y=0) = 2500). 
This indicates a clear violation of scaling with ${\rm N}_{\rm bin}$ 
at the LHC.
These values increase by a factor of 2 to 3 if the effects
of shadowing are not included 
( N$_{\rm ch}$ (y=0) $\approx$ 5000 and 
$\sigma^{\rm sNN}_{{\it c}\,\bar{{\it c}}}$ $\approx$ 8.4 mb).


In summary, we studied the influence of possible strong homogenous 
constant color fields in open charm production 
in heavy-ion collisions by varying the effective string tension that
control Q\=Q pair creation rates. This is equivalent with 
assuming an in medium mass modification of charm quark. 
We show that this approach is an important 
dynamical mechanism that can explain the observed 
{\it D}-mesons enhancement production observed by the PHENIX experiments.
Our model is based on the time-independent color field while in 
reality the production of Q\=Q pairs is a far-from-equilibrium,
time-dependent phenomenon. Thus to achieve more quantitative
conclusions, such mechanisms \cite{gies_jul08}
should be considered in future generation of Monte Carlo codes.

The large cross sections reported by the STAR collaboration 
remain unexplained within our study. 
Solving the discrepancy between the measurements is important, since
confirmation of the STAR results
may indicate the importance of other dynamical mechanisms such 
as pre-equilibrium production 
from secondary parton cascades \cite{wang_prl92}, 
or hot-glue scenario \cite{shuryak_prl92}.

\vskip 0.2cm 
{\bf Acknowledgments:} We thank C. Gale and S. Jeon for useful discussions.
This work was supported by the Natural Sciences and Engineering 
Research Council of Canada.  
This work was supported also by the Division of Nuclear Science, 
of the U. S. Department of Energy under Contract No. DE-AC03-76SF00098 and
DE-FG02-93ER-40764.

\end{document}